\input{aipcheck}

\documentclass[
    ,final            
  ]
  {aipproc}

\layoutstyle{8x11single}

\begin{document}

\title{Center vortices as composites of monopole fluxes}


\author{S.~Deldar }{
  address={Department of Physics, University of Tehran, P.O. Box 14395/547, Tehran 1439955961,
Iran}
}
\author{S.~M.~Hosseini Nejad}{
altaddress={Department of Physics, University of Tehran, P.O. Box 14395/547, Tehran 1439955961,
Iran}
}

\begin{abstract}
 We study the relation between the flux of a center vortex obtained from the center vortex model and the flux formed between monopoles obtained from the Abelian gauge fixing method. Motivated by the Monte Carlo simulations which have shown that almost all monopoles are sitting on the top of vortices, we construct the fluxes of center vortices for $SU(2)$ and $SU(3)$ gauge groups using fractional fluxes of monopoles. Then, we compute the potentials in the fundamental representation induced by center vortices and fractional fluxes of monopoles. We show that by combining the fractional fluxes of monopoles one can produce the center vortex fluxes for $SU(3)$ gauge group in a "center vortex model". Comparing the potentials, we conclude that the fractional fluxes of monopoles attract each other.

\end{abstract}

\maketitle

\section{INTRODUCTION}

  There are numerical evidences supporting the center vortex theory of quark confinement. Furthermore, according to Monte Carlo data, monopoles are also playing the role of agents of confinement. Therefore one can expect some kind of relations between these two topological defects. Lattice simulations by Faber $\it{et~ al.}$ \cite{Del Debbio1998} show that a center vortex configuration after transforming to maximal Abelian gauge and then Abelian projection appears in the 
form of the monopole-vortex chains in $SU(2)$ gauge group. Therefore monopoles and center vortices correlate to each other.
 In $SU(3)$ gauge group, monopoles and vortices form nets where three vortices meet at a single point \cite{Cornwall1998,Chernodub2009}. 
 
 Motivated by these evidences, we construct the fluxes of center vortices for $SU(2)$ and $SU(3)$ gauge groups using fractional fluxes of monopoles. 
We study quark confinement for the fundamental representation in $SU(3)$ gauge group by creating the fractional fluxes of the monopoles, where combinations of them may produce the center vortex fluxes on the minimal area of a Wilson loop in a "center vortex model". Comparing the potential induced by fractional fluxes of monopoles with the one induced by center vortices, it seems that the fractional fluxes attract each other. 

\section{The magnetic charges of Abelian monopoles}
\label{2}
The formation of monopoles is related to the Abelian gauge fixing. The points in space where the Abelian gauge fixing becomes undetermined, are sources of magnetic monopoles. We briefly explain the Abelian gauge fixing method in $SU(2)$ gauge group \cite{Ripka}. A scalar field $\Phi \left( x\right)$ in the adjoint representation is defined as the following:
\begin{equation}
\Phi \left( x\right) =\Phi _a\left( x\right) \mathcal{H}_{a},  \label{phigauge}
\end{equation}
where $\mathcal{H}_{a}$ are the generators of the $SU\left( 2\right)$
group. One can diagonalize the scalar field by a gauge transformation. The gauge in which the scalar field is diagonal is called an Abelian gauge. A degeneracy of the eigenvalues of the scalar field occurs at specific points. The scalar field in the vicinity of the points has hedgehog shape $\it{i.e.}$ $\Phi \left( \vec{r}\right) =x_{a}\mathcal{H}_{a}$. Now, we consider a gauge transformation that diagonalizes the hedgehog field. The gluon field under the same gauge transformation can be separated into a regular part $\vec{A}
^{R} $ and a singular part:
\begin{equation}
\vec{A}=\vec{A}_{a}\mathcal{H}_{a}=\vec{A}_{a}^{R}\mathcal{H}_{a}-\frac{1}{e}\vec{n}_{\varphi }%
\frac{1+\cos \theta }{r\sin \theta }\mathcal{H}_{3},  \label{asph}
\end{equation}
where only the diagonal (Abelian) part of the gluon field
acquires a singular form. The singular part of the gauge field corresponds to the magnetic monopole with the magnetic charge 
\begin{equation}
\label{gpie}
g=-\frac{4\pi }{e}\mathcal{H}_{3}, 
\end{equation}
where $e$ is the color electric charge.

For $SU(3)$ gauge group, the topological defects of
Abelian gauge fixing are sources of magnetic monopoles with magnetic charges equal to:
\begin{eqnarray}
\begin{array}{llll}
g_1=-\frac{4\pi }{e}\mathcal{H}_{3},\\
g_2=-\frac{4\pi }{e}\left( -\frac{1}{2}\mathcal{H}_{3}+\frac{\sqrt{3}}{2}%
\mathcal{H}_{8}\right),\\
g_3=\frac{4\pi }{e}\left( \frac{1}{2}\mathcal{H}_{3}+\frac{\sqrt{3}}{2}%
\mathcal{H}_{8}\right).\\
\end{array} \
\label{g}
\end{eqnarray}
In the next section, we study center vortices in the thick center vortex model.
\section{ Thick center vortex model}
\label{3}
In this model, it is assumed that the effect of a thick center vortex on a fundamental Wilson loop is to multiply the loop by a group factor 
\begin{equation}
\label{group factor}
 \mathcal{G}_f(\alpha^{(n)}) = \frac{1}{d_f}\mathrm{Tr}\left(\exp\left[i\vec{\alpha}^{(n)}\cdot\vec{\mathcal{H}}\right]\right),
\end{equation}
where the $\{\mathcal{H}_i,~i=1,..,N-1\}$ are the Cartan generators and $\vec{\alpha}^n$ shows the flux profile for the center vortex of type $n$. If a thick center vortex is entirely contained within the loop, then
\begin{equation}
\label{center}
\exp\left[i\vec{\alpha}^{(n)}\cdot\vec{\mathcal{H}}\right]=z_n I=e^{\frac{2\pi in}{N}} I~~~~~~~~~~~~~ n=1,2,...,N-1.
\end{equation}
 
 For $SU(2)$ where $z_1=e^{\pi i}$ using Eq. (\ref {center}),
 the maximum value of the flux profile corresponding to the Cartan generator $\mathcal{H}_{3}$ for the fundamental representation is equal to $2\pi$ $\it{i.e.}$
\begin{equation}
\label{center2}
\exp\left[i2\pi \mathcal{H}_3\right]=z_1 I,
\end{equation}
 and for $SU(3)$ where $z_1=e^{\frac{2\pi i}{3}}$, the maximum values of the flux profiles corresponding to the Cartan generators $\mathcal{H}_{3}$ and $\mathcal{H}_{8}$ for the fundamental representation are equal to zero and $\frac{4\pi}{\sqrt{3}}$, respectively. Therefore 

\begin{equation}
\label{center3}
\exp\left[i\frac{4\pi}{\sqrt{3}} \mathcal{H}_8\right]=z_1 I.
\end{equation}
 
The induced potential between static sources in the fundamental representation
 is as the following \cite{Fabe1998}:
\begin{equation}
\label{potential}
V_f(R) = -\sum_{x}\ln\left\{ 1 - \sum^{N-1}_{n=1} f_{n}
(1 - {\mathrm {Re}}\mathcal{G}_{f} [\vec{\alpha}^n_{C}(x)])\right\},
\end{equation}
where $f_n$ is the probability that any given unit area is pierced by a center vortex of type $n$.  
\section{Abelian monopoles and center vortices}
\label{4}
In this section, we construct the center vortex fluxes using fractional fluxes of monopoles. In $SU(2)$ gauge group, when a center vortex completely contains the minimal area of the Wilson loop, Eq. (\ref {center2}) is applied. Substituting $\mathcal{H}_3$ from Eq. (\ref{gpie}) in Eq. ({\ref{center2}) gives
\begin{equation}
\label{cent}
\exp\left[i2\pi \mathcal{H}_3\right]=\exp\left[-ie\frac{g}{2}\right]=z_1 I,
\end{equation}
where $z_1 I=e^{\pi i} I$ is the center element of $SU(2)$ gauge group. In this case, the flux that the center vortex carries is equal to $\pi$: ${\Phi}_v=\pi$. On the other hand, when a closed surface $S$
surrounds a magnetic monopole of charge $g$, the total
magnetic flux crossing the surface is equal to the magnetic charge $g$
of the monopole \cite{Chatterjeea2014}: 
\begin{equation}
\label{ce}
{\Phi}_m=\int_S\vec{B}.d\vec{s}=g
\end{equation}
where $g$ is the monopole charge in Eq. (\ref{gpie}).

Therefore according to Eq. (\ref{cent}) the effect of a center vortex on the Wilson loop is equivalent to the effect of an Abelian configuration related to the half of the matrix flux $g$ on the Wilson loop. The contribution of this Abelian configuration on the Wilson loop is as the following
   \begin{equation}
\label{c8}
W=\mathcal{G}_f=\frac{1}{d_f}\mathrm{Tr}\left(\exp\left[-ie\frac{g}{2}\right]\right)=\frac{1}{2}\mathrm{Tr}\left(\begin{array}{cc} e^{-i\pi} & 0 \\0 &e^{i\pi} 
\end{array} \right)=e^{i\pi}.
\end{equation}  
On the other hand, the contribution
of an Abelian field configuration to the Wilson loop is $W=e^{iq{\Phi}}$ corresponding to $q$ units of the electric charge and for the fundamental representation $q=1$ \cite{Chernodub2005}. 
Therefore the flux of this Abelian configuration corresponding to half of the magnetic charge $g$ which affects on the fundamental Wilson loop is $\pi$. 

As a result, the flux of the Abelian configuration corresponding to half of the magnetic charge $g$ on the Wilson loop is equal to the flux of center vortex on the Wilson loop.

According to the Monte Carlo simulations, almost all monopoles are sitting on top of the vortices by Abelian projection \cite{Del Debbio1998} as shown in Fig. \ref{fig1}. 

\begin{figure}
\includegraphics[scale=0.6]{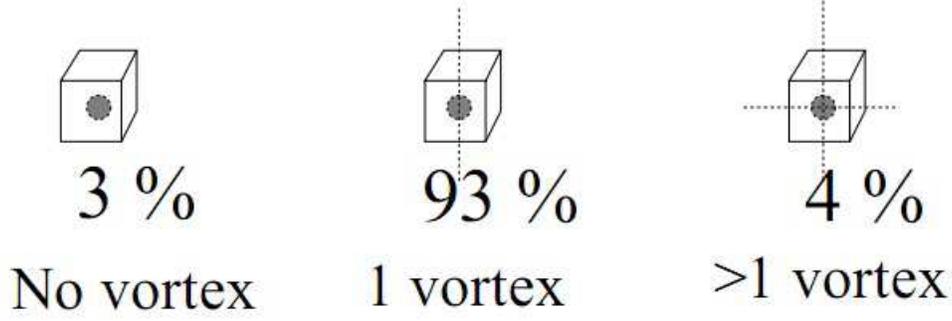}
\caption{Percentages of monopoles pierced by zero, one, or more than one P-vortex line in the Monte Carlo simulations. Almost all of monopoles (about 93\%) are pierced by a single  P-vortex line. Only very small fractions of monopoles either are not pierced 
   at all, or are pierced by more than one line \cite{Del Debbio1998}.}
\label{fig1}
\end{figure}
In addition, the monopole-vortex junctions are also discussed by Cornwall \cite{Cornwall1998} where they are called nexuses. In $SU(N)$ gauge group, each nexus is a source of $N$ vortices which meet each other at a point. 
Using Eq. (\ref{cent}), some configurations which appear in the $SU(2)$ monopole vacuum  
are plotted in Fig. \ref{fig2}. The sum of vortex fluxes must yield the total flux of the monopole. Therefore each monopole 
is a source of two vortex fluxes. A line in the configurations carries half of the flux of the monopole which is equivalent to the center vortex flux. 
\begin{figure}
\includegraphics[scale=0.6]{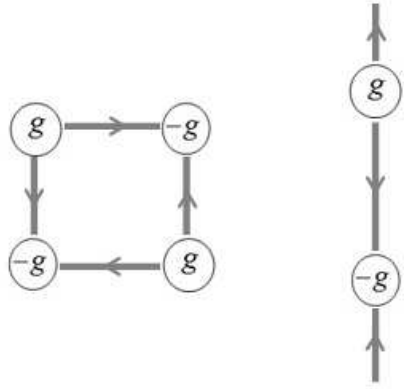}
\caption{Some vortex-monopole configurations in $SU(2)$ vacuum. A line which connects two monopoles carries half of the flux of the monopole.}
\label{fig2}
\end{figure}

In $SU(3)$ gauge group, magnetic charges satisfy the constraint
\begin{equation}
\label{constraint}
g_1+g_2+g_3=0.
\end{equation}
Therefore the number of independent magnetic charges reduces to $2$.
We combine fractional fluxes of magnetic charges to obtain a center vortex flux. When a center vortex pierces the minimal area of the loop, substituting $\mathcal{H}_8$ from Eq. ({\ref{g}) in Eq. (\ref{center3}) gives
\begin{equation}
\label{cen1}
\exp\left[i\frac{4\pi}{\sqrt{3}}\mathcal{H}_8\right]=\exp\left[ie(\frac{g_3}{3}-\frac{g_2}{3})\right]=z_1 I,
\end{equation}
where $z_1 I=e^{\frac{2\pi i}{3}} I$ is the center element of $SU(3)$ gauge group. Therefore, the same as $SU(2)$ gauge group, the effect of a center vortex on the Wilson loop is equivalent to the effect of an Abelian configuration related to one third of the magnetic charge $g_3-g_2$ on the Wilson loop. 

In other words, for $SU(3)$ gauge group, the vortex carries one third of the total monopole flux $g_3$ plus one third of the total monopole flux $g_2$ pointing in opposite direction. On the other hand, using Eq. (\ref{cen1}), some monopole configurations which appear in the $SU(3)$ monopole vacuum are plotted  
 in Fig. \ref{fig3}. A line with one arrow in the configurations carries one third of the total flux of a monopole and a line with two arrows carries two third of the total flux of a monopole.
Since the $SU(3)$ monopole charges satisfy the Dirac quantization condition $eg=2n\pi$, $\exp\left[\pm ieg\right]=1$. Using this equation, two third of the total monopole flux on the Wilson loop may be regarded as the same as one third of the total monopole flux pointing in opposite direction as the following
\begin{equation}
\label{cenn}
\exp\left[ie\frac{2g_n}{3}\right]=\exp\left[ie\frac{2g_n}{3}-ieg_n\right]=\exp\left[ie\frac{-g_n}{3}\right].
\end{equation}
where $g_n$ are monopole charges in Eq. (\ref{g}). Using Dirac quantization condition, right panel of Fig. \ref{fig3} is plotted in Fig. \ref{fig4}. A line with two arrows may be regarded as the same as a line with one arrow pointing in opposite direction. Using Eqs. (\ref{cen1}) and (\ref{cenn}), some configurations which may appear in the $SU(3)$ monopole vacuum  
are plotted in Fig. \ref{fig5}. A line in the configurations corresponds to the center vortex flux which obtains from combining of a line corresponding to one third of the total flux of $g_3$ monopole and a line corresponding to one third of the total flux of $g_2$ monopole pointing in opposite direction.
\begin{figure}
\includegraphics[scale=0.6]{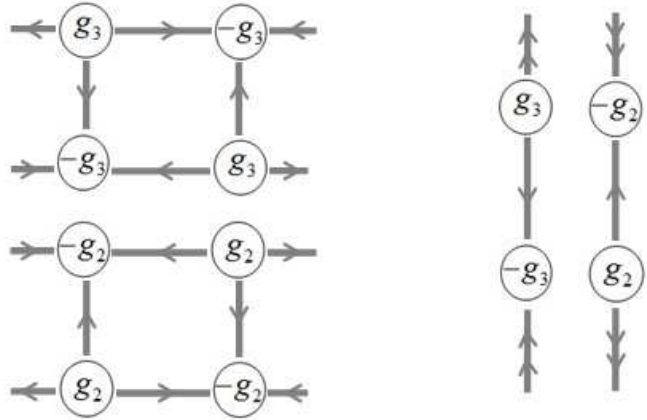}
\caption{Some configurations of the monopole fractional fluxes in $SU(3)$ vacuum. A line with one arrow corresponds to one third of the total monopole flux and a line with two arrows corresponds to two third of the total monopole flux.}
\label{fig3}
\end{figure}
 \begin{figure}
\includegraphics[scale=0.6]{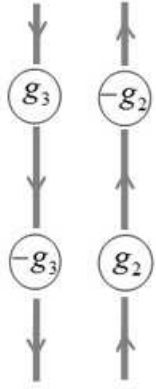}
\caption{Same as right panel of Fig. \ref{fig3}. Using Dirac quantization condition, two third of the total monopole flux on the Wilson loop may be regarded as the same as one third of the total monopole flux pointing in opposite direction.}
\label{fig4}
\end{figure}
\begin{figure}
\includegraphics[scale=0.6]{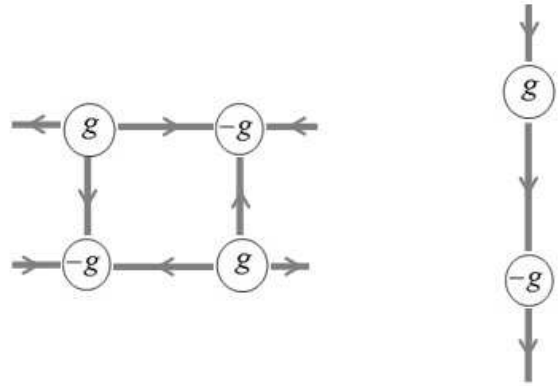}
\caption{Some vortex-monopole configurations in $SU(3)$ vacuum. The lines in the configurations correspond to the center vortices and the magnetic charge $g$ is equal to $g_3-g_2$.}
\label{fig5}
\end{figure}
Now, we investigate the potentials induced by fractional flux lines of the monopoles in a "center vortex model". We use quotation mark because it is different from the center vortex model. The difference comes from the fact that we use each fractional flux between monopoles as an individual "vortex".
\section{Potential for fundamental representation}
 \label{5}
In the center vortex model, the center vortices are line-like and the Wilson loop is considered a rectangular $R\times T$ loop
in the $x-t$ plane, with $T>>R$.  The 
time-extension $T$ is huge but fixed. Therefore the loop is characterized  
just by the width $R$. 

 Among vortex-monopole configurations, configurations which are line-like (right configurations in Figs. \ref{fig2} and \ref{fig5}) affect the Wilson loop. Other configurations have no effect on the Wilson loop because the effect of a center vortex inside a configuration is eliminated by another center vortex pointing in opposite direction. 

As stated above, in $SU(3)$ gauge group, the combination of the fractional fluxes of monopoles leads to the center vortex flux. Now, the interaction between these fractional fluxes is investigated in a "center vortex model". 

Creation of the fractional flux line of $SU(3)$ monopole vacuum linked to the fundamental representation Wilson loop in a "center vortex model" has the effect of multiplying the Wilson loop by a phase, $\it{i.e.}$
\begin{equation}
\label{cenv}
W_f(C) \to~ \mathcal{G}_f W_f(C),
\end{equation}
where $\mathcal{G}_f =\frac{1}{d_f}\mathrm{Tr}e^{ie\frac{g_n}{3}}$ and $g_n$ $(n=2,3)$ are monopole charges in Eq. (\ref{g}).
 The static potential induced by fractional flux lines of the monopoles (see Fig. \ref{fig4}) in this "center vortex model", where combinations of them produce the center 
vortex fluxes is as the following:
\begin{equation}
\label{potential}
V_f(R) = -\sum_{x}\ln\left\{ 1 - \sum^{3}_{n=2} f_{n}
[1 - {\mathrm {Re}}(\mathcal{G}_{f})]\right\}.
\end{equation}
On the other hand, the static potential induced by thin center vortices is 
\begin{equation}
\label{potential1}
V_f(R) = -\sum_{x}\ln\left\{ 1 -  f_{1}
[1 - {\mathrm {Re}}(z_1)]\right\}.
\end{equation}
Figure \ref{fig6} shows the potentials induced by fractional fluxes of the monopoles and center vortices for the fundamental
 representation.
 
 \begin{figure}
\includegraphics[scale=0.5]{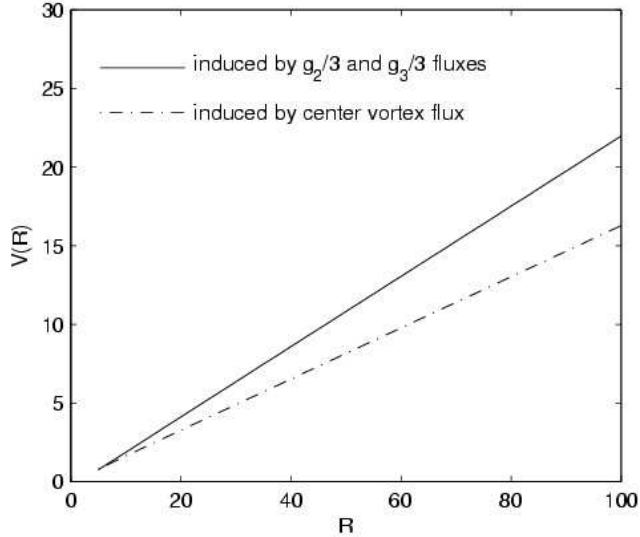}
\caption{Potential $V(R)$ induced by monopole fractional fluxes where combinations of them produce the center 
vortex fluxes as well as the one induced by center vortices in $SU(3)$ gauge theory. Upper linear potential is induced by an interaction between the Wilson loop and fractional fluxes corresponding to one third of $g_3$ monopole fluxes and one third of $g_2$ monopole fluxes. The probability that any given plaquette is pierced by a
fractional flux is chosen to be $0.1$. Lower linear potential is induced by an interaction between the Wilson loop and center vortex fluxes. The probability that any given plaquette is pierced by a
center vortex flux is chosen to be $0.1$. The potential induced by center vortices has an extra negative energy compared to the one induced by fractional fluxes where combinations of them produce the center vortex fluxes. Therefore it may be an attraction between fractional flux lines in Fig. \ref{fig4} to produce center vortex line.}
\label{fig6}
\end{figure}
The free parameters $f_n$ in Eqs. (\ref{potential}) and (\ref{potential1}) are chosen to be $0.1$. As shown in Fig. \ref{fig6}, the potential induced by fractional fluxes of monopoles is linear. 
The extra negative potential energy of static potential induced by center vortex compared with the potential induced by fractional fluxes of monopoles shows that the fractional flux lines of the monopoles in Fig. \ref{fig4}, attract each other. 
\section{conclusion}
 \label{6}
Both the Abelian monopole and center vortex mechanisms of the quark confinement are supported by
lattice gauge theory. Therefore one can expect that Abelian monopoles are related to center vortices. According to lattice results in $SU(2)$ gauge theory, a center vortex configuration, 
transformed to maximal Abelian gauge and then Abelian-projected, will appear as a 
 monopole-vortex chain. Motivated by these evidences, we construct center vortex fluxes using fractional fluxes of monopoles for $SU(2)$ and $SU(3)$ gauge groups. A "center vortex model" is applied to $SU(3)$ gauge group to calculate induced potentials from the center vortices and the monopole fractional fluxes where combinations of them produce center 
vortex fluxes. Comparing the potentials, we conclude that the fractional fluxes obtained from monopoles attract each other. 

\begin{theacknowledgments}
   We would like to thank M. Faber for the very useful discussions. We are grateful to the research council of the University of Tehran for
supporting this study.
\end{theacknowledgments}

\bibliographystyle{aipproc}

\end{document}